\newif \ifproofs
\proofstrue

\documentclass[11pt]{article}
\usepackage{amssymb}
\usepackage{mathrsfs,amsmath}
\usepackage{graphicx}
\usepackage{amsmath}
\usepackage{amsthm}
\usepackage{bbm}
\usepackage{dsfont}
\usepackage{listing}
\usepackage{hyperref}
\usepackage{color}
\usepackage[ruled, vlined,linesnumbered, noend]{algorithm2e}

\newtheorem{theorem}{Theorem}

\newtheorem{lemma}{Lemma}
\newtheorem{observation}{Observation}

\newtheorem{fact}{Fact}
\newtheorem{assumption}{Assumption}
\theoremstyle{definition}
\newtheorem{definition}{Definition}
\newtheorem{remark}{Remark}
\newtheorem{example}{Example}
\allowdisplaybreaks

\newcommand{\abs}[1]{\left| #1 \right|}
\newcommand{\bigpar}[1]{\left( #1 \right)}
\newcommand{\bigbra}[1]{\left[ #1 \right]}
\newcommand{\bigbrace}[1]{\left\{ #1 \right\}}

\newcommand{\casewise}[1]{\left\{ #1 \right.}

\newcommand{\cD}{\mathcal{D}}

\newcommand{\cL}{\mathcal{L}}

\newcommand{\cX}{\mathcal{X}}

\usepackage{fancyhdr}
\usepackage{calc}
\fancyheadoffset[RE]{\marginparsep+\marginparwidth}

\usepackage[top=1.5in, bottom=1in, left=1in, right=1in]{geometry}

\graphicspath{ {peripherals/fig/} }

\title{Private Location Sharing for Decentralized Routing services}

\author{

\begin{tabular}{cc}
    \begin{tabular}{c}
        Matthew Tsao \\
        Stanford University \\
        \texttt{mwtsao@stanford.edu} 
    \end{tabular}
     & 
    \begin{tabular}{c}
        Kaidi Yang \\
        Stanford University \\
        \texttt{kaidi.yang@stanford.edu} 
    \end{tabular}
    \\
    & \\
    \begin{tabular}{c}
        Karthik Gopalakrishnan \\
        Stanford University \\
        \texttt{gkarthik@stanford.edu} 
    \end{tabular}
     & 
    \begin{tabular}{c}
        Marco Pavone \\
        Stanford University \\
        \texttt{pavone@stanford.edu} 
    \end{tabular}    
\end{tabular}

}

\begin{document}
\maketitle

\begin{abstract}
    Data-driven methodologies offer many exciting upsides, but they also introduce new challenges, particularly in the realm of user privacy. Specifically, the way data is collected can pose privacy risks to end users. In many routing services, a single entity (e.g., the routing service provider) collects and manages user trajectory data. When it comes to user privacy, these systems have a central point of failure since users have to trust that this entity will not sell or use their data to infer sensitive private information. Unfortunately, in practice many advertising companies offer to buy such data for the sake of targeted advertisements.
    
    With this as motivation, we study the problem of using location data for routing services in a privacy-preserving way. Rather than having users report their location to a central operator, we present a protocol in which users participate in a decentralized and privacy-preserving computation to estimate travel times for the roads in the network in a way that no individuals' location is ever observed by any other party. The protocol uses the Laplace mechanism in conjunction with secure multi-party computation to ensure that it is cryptogrpahically secure and that its output is differentially private. 
    
    A natural question is if privacy necessitates degradation in accuracy or system performance. We show that if a road has sufficiently high capacity, then the travel time estimated by our protocol is provably close to the ground truth travel time. We validate the protocol through numerical experiments which show that using the protocol as a routing service provides privacy guarantees with minimal overhead to user travel time.
\end{abstract}

\section{Introduction}\label{sec:intro}

Big Data and data-driven methodologies have shown promise in improving the efficiency, safety and adaptability of mobility services. However, certain types of data sharing can also lead to privacy risks for users. In this paper we focus on merits and risks of sharing location data. We discuss how location data is useful for determining congestion levels in routing services (e.g., Google Maps, Apple Maps, Waze), and we discuss user privacy risks involved with location sharing. With this as motivation we show how a protocol for decentralized location sharing can mitigate privacy risks while retaining some of the merits of location information for routing services.

\begin{figure}
    \centering
    \includegraphics[width = 0.75\textwidth]{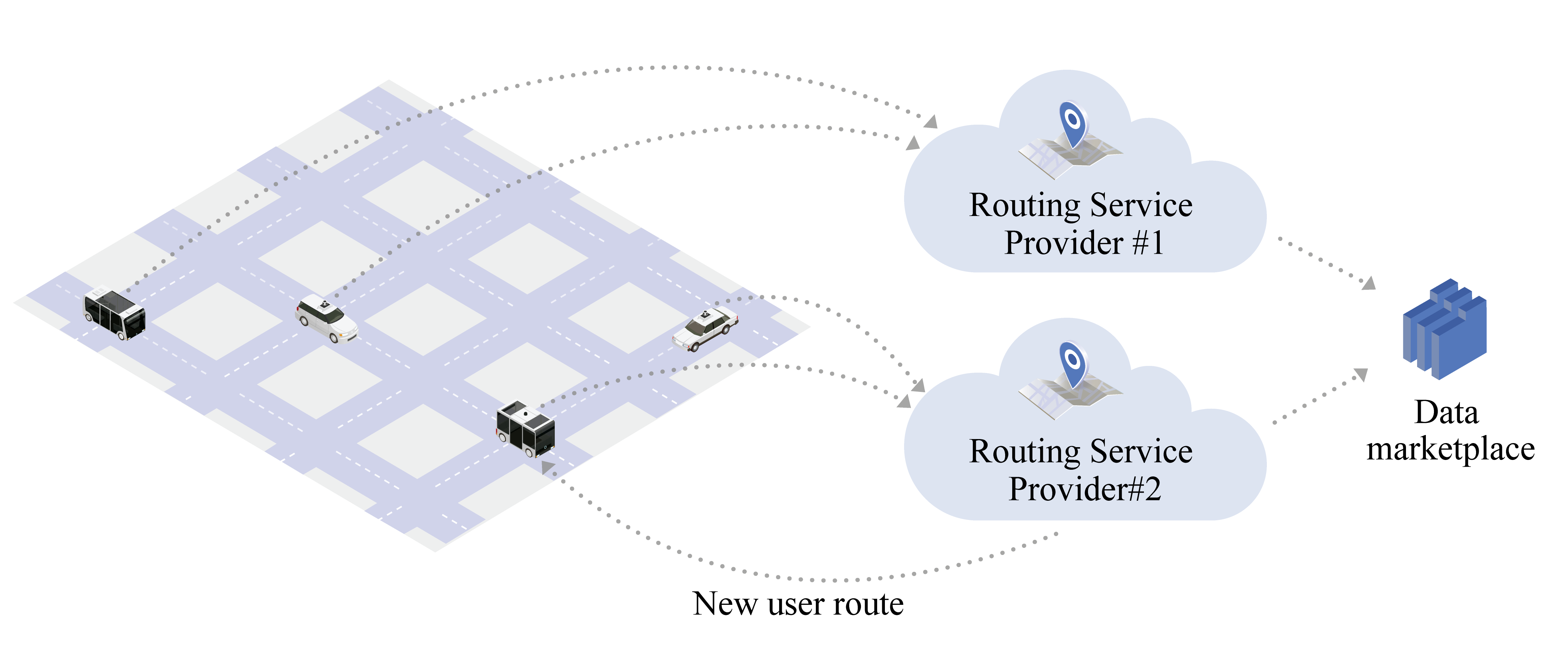}
    \caption{In most routing services, users give their location data in exchange for route recommendations. Routing services often sell this data in data marketplaces. Third parties who buy location data from these routing services will be able to infer preferences, habits, and schedules of users who frequently interact with routing services.
    }
    \label{fig:centralized}
\end{figure}

Repeated exposure to conventional location sharing can lead to privacy risks for users. In many current routing services, users provide their location data in exchange for routing recommendations. While users often only provide a small amount of their location data each time that they use a routing service, if a user regularly uses routing services, the data they share over many interactions can be stitched together to form a more complete picture of the user's routines, behaviors, preferences, etc. User privacy in such settings thus requires trust that the routing services will not share user data with other entities. However in practice, advertising companies offer to buy this user data to build user profiles for the sake of targeted advertising. As a result, even though users only share small amounts of their location data in each interaction with a routing service, a single entity may end up with a large amount (likely more than the user is comfortable with) of their location data (see Fig.~\ref{fig:centralized}).

While location sharing presents privacy challenges, it also provides utility for routing services. Location information is helpful because congestion levels of a road can be estimated from the number of vehicles on the road. A key insight toward addressing privacy challenges is that the congestion level only depends on aggregate location information; what matters is the \textit{number} of vehicles on a road, not which particular users are on the road. This suggests that aggregation procedures can be used to protect individual user location while still providing the location information needed for routing services. 

\subsection{Statement of Contributions}\label{sec:intro:soc}
Motivated by this observation, in this paper we propose a decentralized location sharing protocol where users on the road will periodically compute and announce the traffic counts (e.g., approximate number of vehicles traveling on each road) of the transportation network in a decentralized and privacy-preserving manner. Since only the total number of vehicles on each road is announced, the location of individual users is not discernible by observers, which is contrary to many current location sharing setups where users give their individual location data directly to routing services. With this protocol, user privacy does not rely on a trusted data custodian, and there are no single points of failure. 

\begin{figure}
    \centering
    \includegraphics[width = 0.75\textwidth]{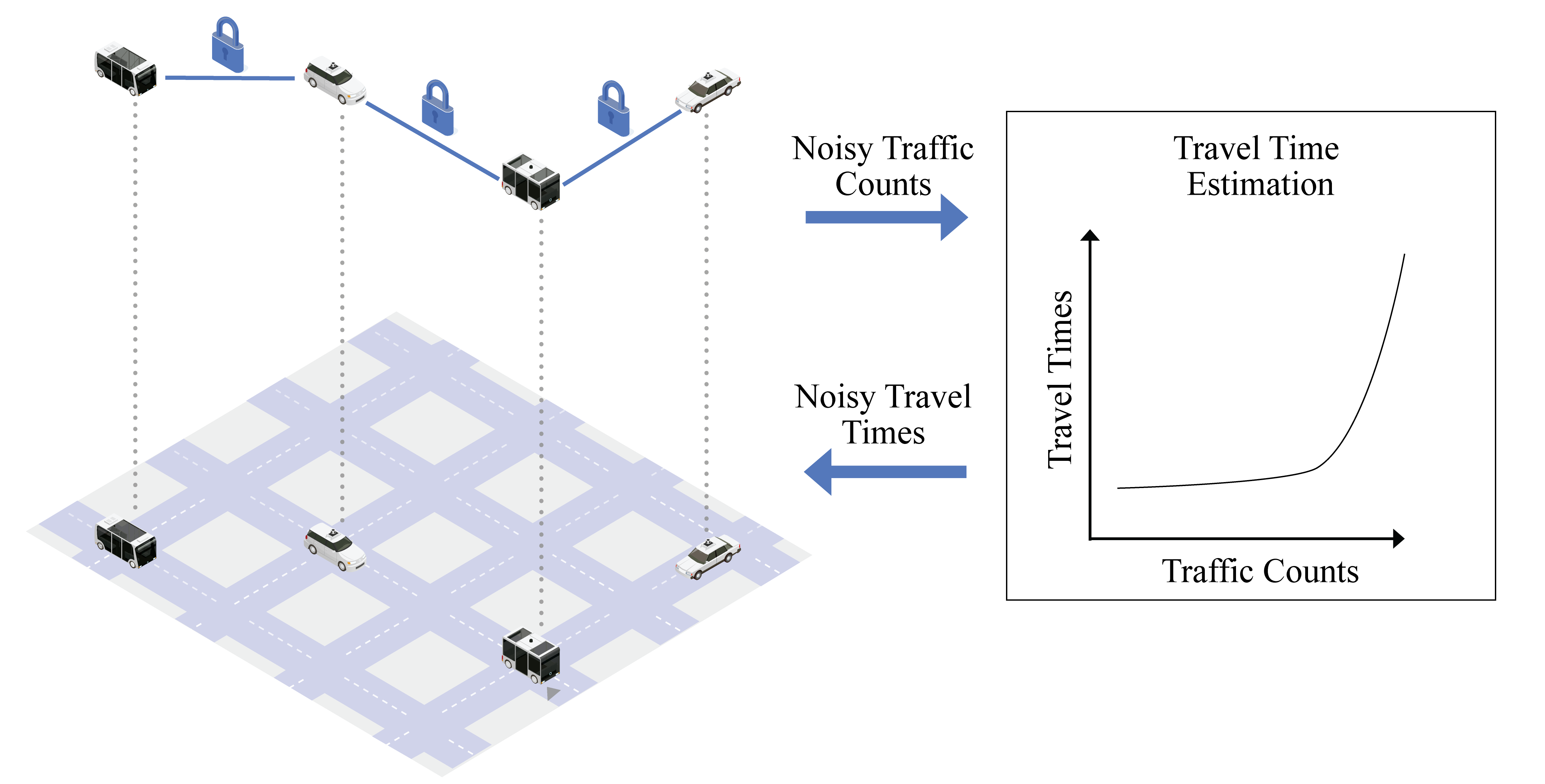}
    \caption{A visualization of the routing service protocol described in Algorithm~\ref{alg:main}. Users traveling in the transportation network share their location data in a  privacy-preserving way to estimate the traffic counts in a decentralized manner (upper left). These counts are then used to estimate travel times (right). When a user requests a route from the routing service, a shortest path is computed using the estimated travel times (lower left).}
    \label{fig:protocol}
\end{figure}

Furthermore, assuming the roads in the network are sufficiently large, we can prove that the travel time estimates produced by the protocol will be close to the estimates produced by the ground truth with high probability. This result showcases an interesting complementarity between differential privacy and delay functions used in travel time estimation. In low traffic situations, differential privacy constraints lead to poor accuracy for traffic count estimation. However, delay functions are insensitive for small inputs and can thus tolerate the poor accuracy. On the other hand, delay functions are very sensitive in high traffic situations, and differential privacy can provide high accuracy in these settings. Thus when a delay function is composed with a differentially private mechanism, the two compensate for the others' weaknesses to yield accurate and private travel time estimates. We corroborate this insight using numerical experiments which show that the protocol provides a privacy-preserving routing service with minimal overhead to the travel time of users.

\subsection{Related Work}\label{sec:intro:related_work}

Privacy research in transportation mainly focuses on \emph{location privacy}, whereby the aim is to prevent untrusted entities from learning geographic locations or location sequences of an individual \cite{BeresfordStajano2003}. 
A number of privacy-preserving approaches have been proposed for various location-based applications, e.g.,  trajectory publishing, mobile crowdsensing, traffic control, etc. 
From a methodological perspective, these approaches are often implemented through spatial cloaking~\cite{ChowMokbelEtAl2011}, differential privacy\cite{Dwork2008}, and Secure Multi-Party Computation (MPC) \cite{GoldreichMW87}. 

Spatial cloaking-based approaches rely on aggregation to convert users' exact locations to coarse information. These approaches are often based on k-anonymity \cite{Sweeney2002}, where a mobility dataset is divided into equivalence classes based on data attributes (e.g., geological regions, time, etc.) so that each class contains at least k records \cite{GhasemzadehFungEtAl2014,HeChow2020}. These k-anonymity-based approaches can guarantee that every record in the dataset is indistinguishable from at least k-1 other records. However, k-anonymity is generally considered to be a weak privacy guarantee. Furthermore, due to coarse data aggregation, spatial cloaking-based approaches can lead to low data accuracy. 

Differential privacy-based approaches provide a sound privacy guarantee by producing randomized responses to queries, whereby two datasets that differ in only one entry produce statistically indistinguishable responses \cite{DworkMNS06}.
In other words, differential privacy ensures that an adversary with arbitrary background information (e.g., query responses, other entries) cannot infer individual entries with high confidence. 
Existing research for location data either probabilistically generates obfuscated locations from a user's true location \cite{WangHuEtAl2018,YanLuEtAl2019} or adds noises to the number of users within each equivalent class \cite{GongZhangEtAl2015,GursoyLiuEtAl2018,HussaeniFungEtAl2018,DongKricheneEtAl2015,LiYangEtAl2020}. However, differential privacy-based approaches can suffer from two drawbacks. First, due to randomization, there is a trade-off between the accuracy of the response and the level of privacy. Second, most existing research requires a trusted data collector to generate random responses, which does not fit our decentralized setting in this paper. 

Secure MPC serves as an excellent technique for decentralized settings, whereby several players jointly compute a function over their data while keeping these data private. Existing secure MPC-based research proposes traffic monitoring and control approaches that keep users' location data confidential, based on secret sharing \cite{WernkeDurrEtAl2013,RenTang2020}, homomorphic encryption \cite{LiLinEtAl2019,ZhangYangEtAl2020}, and blockchain \cite{ZouXiEtAl2021}. Secure MPC can ensure accuracy since no noises are added to protect location privacy. However, Secure MPC can suffer from high computational overhead due to encryption, and the computation results might leak private information (See Remark~\ref{rem:mpc_leak} for more details). 

\subsection{Organization}

This paper is organized as follows. In Section~\ref{sec:model} we present a model for the transportation system, and specify both the system objective and privacy requirements. We present a decentralized and privacy-preserving routing service protocol in Section~\ref{sec:method} along with all of the statistical and cryptographic tools used by the protocol. In Section~\ref{sec:theory} we prove that if the roads in the transportation network are sufficiently large, then the protocol provides a privacy-preserving routing service whose travel time estimates are provably close to the ground truth. We evaluate our protocol in numerical experiments and present the results in Section~\ref{sec:experiments}. We summarize our work and identify important areas for future work in Section~\ref{sec:conclusion}. 

\section{Model}\label{sec:model}

In this section we describe the transportation network model, the objective for the users' distributed algorithm to estimate traffic counts, and the privacy requirements for the algorithm.

\subsection{Transportation Network}\label{sec:model:transit_network}

The transportation network is represented as a directed graph $G := (V, E)$ where edges $E$ represent roads and vertices $V$ represent road intersections. We use $n := \abs{V}$ and $m := \abs{E}$ to denote the number of vertices and edges in the graph respectively. The concepts of traffic flow, traffic counts, and travel times are essential to this work, so we will describe them here. 

\begin{definition}[Traffic Flow]\label{def:traffic_flow}
For a given road $e \in E$, its traffic flow $x_e$ measures the number of vehicles that enter the road during a fixed time interval (e.g., every second). 
\end{definition}

\begin{definition}[Travel Time]\label{def:travel_time}
Each edge $e \in E$ has an associated delay function $f_e : \mathbb{R} \rightarrow \mathbb{R}$ where $f_e(x_e)$ is the estimated travel time on the road $e$ if the traffic flow on the edge is $x_e$. 
\end{definition} 

\begin{definition}[Traffic Counts]\label{def:traffic_counts}
For a given road $e \in E$, its traffic count $s_e$ is the number of vehicles currently on the road. At steady state the traffic count is equal to the traffic flow multiplied by the travel time. Specifically, $s_e = x_e f_e(x_e)$. For convenience, we define the flow-counts function $F_e$ as $F_e(x_e) := x_e F_e(x_e)$ so that $s_e = F_e(x_e)$.
\end{definition}

\noindent Throughout this paper we make the following natural assumption on delay functions.

\begin{assumption}[Properties of Delay Functions]\label{assump:traveltime}
We assume that for each road $e \in E$, $f_e$ is a positive, non-decreasing and differentiable function on $\mathbb{R}_+$. 
\end{assumption}

\begin{remark}\label{rem:bpr}
The Bureau of Public Roads (BPR) function $f_{\text{BPR}, e}(x_e) := 1 + 0.15 \bigpar{ \frac{x_e}{c_e} }^4$ is a commonly used volume delay function which satisfies Assumption~\ref{assump:traveltime}. Namely, it is a degree $4$ polynomial with positive coefficients (i.e., $c_e > 0$).
\end{remark}

\begin{definition}[Travel Time as a Function of Traffic Counts]\label{def:counts2time}
For each road $e \in E$ we define $\tau_e$ as the function that estimates travel time based on traffic counts. In other words, for an edge $e$ with volume $x_e$ and counts $s_e$, we have $\tau_e(s_e) = f_e(x_e)$. Since we know from Definition~\ref{def:traffic_counts} that $x_e = F_e^{-1}(s_e)$, we have $\tau_e(s_e) := f_e(F_e^{-1}(s_e))$.

\end{definition}

\noindent Road capacities are a concept that will be important to our methodology and results, which we define as follows: 

\begin{definition}[$\delta$-capacity]\label{def:delta_cap}
For $\delta > 0$, the $\delta$-capacity of a road $e$, denoted $c_{e,\delta}$, is the largest value so that for all $x_e \leq c_{e,\delta}$ we have $f_e(x_e) \leq (1+\delta) f_e(0)$. 
\end{definition}

\subsection{Users and Traffic Counts}\label{sec:model:state}

At any given time $t$, let $N(t)$ denote the number of users currently traveling in the transportation network. For $1 \leq i \leq N(t)$, the state of user $i$ at time $t$, given by $s(t,i) \in \bigbrace{0,1}^m$, specifies which road the user is on. The $e^{th}$ entry of the vector $s(t,i)$ is given by:
\begin{align*}
    s_e(t,i) = \mathds{1} \bigbra{\text{user } i \text{ is on road } e}.
\end{align*}
Note that exactly one entry of $s(t,i)$ is $1$ and all others are $0$. The traffic counts at time $t$, denoted $s(t) \in \mathbb{N}^m$, represents the total number of vehicles on each road and is defined as
\begin{align*}
    s(t) := \sum_{i=1}^{N(t)} s(t,i). 
\end{align*}
The number of users traveling on road $e$ at time $t$, denoted by $s_e(t)$, is defined as $s_e(t) = \sum_{i=1}^{N(t)} s_{e}(t,i)$.

\subsection{Communication Model}\label{sec:model:communication}

In this work we assume that users can communicate with one another through private channels. Concretely, this means that for any pair of users $i$ and $j$, user $i$ can send a message that can only be deciphered by user $j$. Such a communication channel can be easily established using standard public key cryptography systems. 


\subsection{System Objective}\label{sec:model:system_objective}

The goal of the system is to periodically broadcast estimated travel times for all roads in the network for the sake of route recommendation. The accuracy of travel time estimates will be measured by mean absolute percentage error (MAPE) as defined in Definition~\ref{def:MAPE}.  

\begin{definition}[Mean Absolute Percentage Error (MAPE)]\label{def:MAPE}
Suppose $T$ is a (possibly randomized) estimator for a positive target value $t^*$. Then the mean absolute percentage error (MAPE) of $T$ is given by
\begin{align*}
    \mathbb{E}_T \bigbra{ \frac{\abs{T - t^*}}{t^*} }
\end{align*}
where the expectation is taken over the randomness in $T$. 
\end{definition}

\noindent The following remark explains how the traffic counts are valuable to this effort.

\begin{remark}[routing service from traffic counts]\label{rem:tt_2_routing}
The functions $\bigbrace{\tau_e}_{e \in E}$ from Definition~\ref{def:counts2time} can be used to compute estimated travel times $\bigbrace{\tau_e(s_e(t))}_{e \in E}$ for all roads at time $t$ from the traffic counts. A routing service can then recommend routes to users based on shortest paths computed from the estimated travel times.
\end{remark}

With Remark~\ref{rem:tt_2_routing} in mind, the system's goal is to compute and announce the traffic counts of the system every $\Delta t$ minutes. Concretely, for each $k \in \mathbb{N}$, at time $k \Delta t$ the $N(k \Delta t)$ traveling users must compute an approximation to $s(k \Delta t)$ in a distributed and privacy-preserving way where privacy is defined according to Definition~\ref{def:privacy}. 

\begin{definition}[Privacy-Preserving Mechanism]\label{def:privacy}
A mechanism is $\epsilon$-\textit{privacy preserving} if it is $\epsilon$-\textit{differentially private} and can be computed in a distributed setting in a way that is \textit{cryptographically secure} against \textit{semi-honest adversaries}. 
\end{definition}

\noindent The precise definitions for cryptographic security, semi-honest adversaries and differential privacy are presented and motivated in the next section. 

\begin{remark}[On the choice of location data for travel time estimation]\label{rem:x_vs_v}
In this work we use location data to estimate travel times in a transportation network. This is done by first estimating the traffic flow from traffic counts, and then estimating travel time from traffic flow. One natural alternative is to have users share both their location and speed. In this alternative approach, the location would specify which road the user is on and the average speed reported on a road could be used to estimate its travel time. We opted not to use speed information for two main reasons. The first is due to privacy requirements. As we will discuss and motivate in Section~\ref{sec:model:privacy:diffpriv}, differential privacy is an important property that we want our method to have. Due to the properties of differential privacy, there are effective ways to compute counts (such as the number of users on a given road) but no clear way to compute an average of user data (such as average speed) in a differentially private way. See Remark~\ref{rem:diffpriv:sensitivity} for more details. The second is for ease of deployment. Requiring only location data means that our protocol only needs sparse GPS measurements, whereas speed estimation needs continuous GPS measurements, which essentially means that users are being tracked. 
\end{remark}

\subsection{Privacy Requirements}\label{sec:model:privacy}

To ensure user privacy, there are two requirements we impose on a desired protocol for the computation of traffic counts: cryptographic security and differential privacy. 

\subsubsection{Cryptographic Security}\label{sec:model:privacy:crypto}

Cryptographic security pertains to settings where a group of agents, each with private data, would like to compute a joint function of everyone's data without any agent needing to reveal its private data to other agents. In our setting, at time $k \Delta t$ the $N(k \Delta t)$ users traveling in the network are agents, where $s(t,i)$ is the private data of the $i$th user, and the desired function is the sum of everyone's private data. We make the following standard assumption on user behavior:

\begin{assumption}[Semi-honest users]\label{assump:semihonest}
We assume that all users are semi-honest\footnote{Semi-honest adversaries, honest-but-curious adversaries, and passive adversaries are equivalent and used interchangeably in the cryptography literature.}, which means they will follow the protocol but may try to do additional computation to learn the secret data of other users. 
\end{assumption}

The definition of cryptographic privacy measures privacy by comparing protocols to an ideal computation model which is defined below. 

\begin{definition}[Ideal Computation Model]
In the Ideal Computation Model, there are $n$ agents $a_1,...,a_n$ with private data $x_1,...,x_n$ wanting to compute $f(x_1,..,x_n)$. Each agent sends its private data to a trusted third party which uses the private data to compute $f(x_1,...,x_n)$ and sends this value back to all of the agents. 
\end{definition}

However, since trusted third parties cannot be assumed to exist, the ideal computation model cannot be implemented in a trustless and decentralized setting. Still, this model serves as a gold standard, and cryptographically secure protocols are required to provide the same level of security as this ideal model.  

\begin{definition}[Cryptographic Security]
A protocol between $n$ agents $a_1,...,a_n$ with private data $x_1,...,x_n$ wanting to compute $f(x_1,..,x_n)$ is cryptographically secure if no probabilistic polynomial time agent learns anything more about other agents' data than they would have learned in the Ideal Computation Model. 
\end{definition}

In other words, a protocol is cryptographically secure if no computationally efficient agent learns more from interacting with the protocol than they would from interacting with the Ideal Computation Model. 

\begin{remark}\label{rem:mpc_leak}
We emphasize that this does not mean that agents learn nothing about other agents' data. This is illustrated by a simple three agent example $a_1, a_2, a_3$ with private data $x_1,x_2,x_3$ and query function $f(x_1,x_2,x_3) = x_1 + x_2 + x_3$. In this example, by learning $f(x_1,x_2,x_3)$, $a_1$ learns the sum of the other agents' data: $x_2 + x_3 = f(x_1,x_2,x_3) - x_1$. 
\end{remark}

In light of Remark~\ref{rem:mpc_leak}, it is more accurate to say cryptographically secure protocols reveal nothing about other agents' data \textit{beyond the value of the output}. 

Cryptographic security is necessary for user privacy, since we certainly do not want users to be able to determine the location of certain individuals through our protocol. 

Unfortunately, for the application of user location data, cryptographic security alone is not enough to ensure user privacy, which we illustrate in the following example. 

\begin{example}[Insufficiency of Cryptographic Privacy for Sparse Data]\label{ex:cryptosec_not_enough}
If Alice is an early bird and wakes up to run errands in the city before anyone else wakes up, then in the morning $N(t) = 1$ since Alice is the only person in the network, and therefore we have $s(t) = s(t, \text{Alice})$, hence the traffic counts reveal Alice's location information. While $s(t)$ does not explicitly label the single traveler in the system as Alice, this information can be inferred if the traveler begins and ends its route at Alice's house. More generally, cryptographic security does not provide user privacy in sparse data settings. Even when there are multiple users active in the network, side information attacks can be used to associate trajectories in sparse datasets to certain individuals \cite{Pandurangan14}. 
\end{example}

This motivates the second privacy requirement we enforce in this paper, which is differential privacy. 

\subsubsection{Differential Privacy}\label{sec:model:privacy:diffpriv}

With Example~\ref{ex:cryptosec_not_enough} in mind, to protect the privacy of users like Alice, the output of a privacy-preserving protocol should not depend too much on the data of any single user. One way to ensure this is through differential privacy. To quantify the influence of a single user, we first introduce the concept of adjacent datasets.

\begin{definition}[Adjacent Datasets]
Two datasets $D_1, D_2$ are adjacent if $D_1$ contains at most one datapoint that is not in $D_2$ and $D_2$ contains at most one datapoint that is not in $D_1$. Concretely, $D_1,D_2$ are adjacent if $\abs{D_1 \setminus D_2} \leq 1$ and $\abs{D_2 \setminus D_1} \leq 1$.
\end{definition}

\noindent In our setting, a dataset $D_1 = \bigbrace{s(t,i)}_{i=1}^{N(t)}$ would be the locations of the $N(t)$ users who are traveling within the transit network at time $t$. The dataset $D_2$ obtained from $D_1$ by modifying the location of one user, who we will call Alice, would be adjacent to $D_1$ since $D_1 \setminus D_2$ contains only the datapoint corresponding to Alice's original location, and $D_2 \setminus D_1$ contains only Alice's newly modified location. One sufficient way to ensure that a mechanism does not depend too much on any single users' data is to demand that the mechanism behaves similarly on adjacent datasets. This is the approach taken by differential privacy which is defined below. 

\begin{definition}[Differntially Private Mechanism]
For $\epsilon > 0$, a $\epsilon$-differentially private mechanism $M : \cD \rightarrow \cX$ is a randomized function mapping datasets into an output space $\cX$ so that for any event $E \subset \cX$ and any adjacent datasets $D_1,D_2$, we have
\begin{align*}
    \mathbb{P}\bigbra{ M(D_1) \in E } &\leq e^\epsilon \mathbb{P}\bigbra{ M(D_2) \in E }
\end{align*}
\end{definition}

\noindent To understand why differential privacy gives us the desired privacy we seek, first note that for any two adjacent datasets $D_1,D_2$, the distributions of $M(D_1), M(D_2)$ are very similar. More specifically, the total variation distance between the distributions of $M(D_1), M(D_2)$ is at most $\epsilon$. Because of this, no hypothesis test can determine from the output of the mechanism whether its input was $D_1$ or $D_2$ with success probability better than $\frac{1 + \epsilon}{2}$, which is barely better than random guessing for small $\epsilon$. This result holds even if the hypothesis test is given knowledge of all datapoints in $D_1 \cap D_2$.  

Now suppose $D_1$ is a dataset that contains Alice's location, and $D_2$ is obtained from $D_1$ by modifying Alice's location arbitrarily. If an observer were able to accurately infer Alice's location based on the output of a $\epsilon$-differentially private mechanism, then it would be able to reliably distinguish between the inputs $D_1$ and $D_2$. However, since differential privacy makes such a task statistically impossible, by contraposition it is statistically impossible for an observer to accurately infer Alice's location based on the mechanism's output. Hence differential privacy ensures privacy of Alice's data. 

The following remark describes a general methodology for achieving differential privacy. 

\begin{remark}[Query sensitivity and the required noise level]\label{rem:diffpriv:sensitivity}
Dwork's pioneering work \cite{DworkMNS06} proposes adding noise to queries in order to achieve differential privacy. Given a data set $D$ and a query $f$, the mechanism $D \mapsto f(D) + Z$ is differentially private so long as $Z$ is a random variable with sufficiently large variance. Specifically, to achieve $\epsilon$-differential privacy, the variance of $Z$ should be at least $\frac{L_f^2}{\epsilon^2}$ where $L_f$ is the sensitivity of the function $f$, which is defined as
\begin{align*}
    L_f := \sup_{\text{Adjacent datasets } D_1,D_2} f(D_1) - f(D_2).
\end{align*}
Revisiting Remark~\ref{rem:x_vs_v}, the role of sensitivity in differential privacy is a main reason why we chose to estimate travel time using counts rather than with average speed. The sensitivity of counting functions is $1$ since the modification of a single data point can change the count by at most $1$, however the sensitivity of an average is unbounded since a large change to a single data point in a data set can lead to a large change in the average. As such, counting functions are much more compatible with the concept of differential privacy than averages are. 
\end{remark}

\section{Methodology}\label{sec:method}

In this section we describe our distributed protocol to enable users to approximately compute $s(t)$ in a privacy-preserving way. We will be using a Laplace Mechanism, which is described in Section~\ref{sec:method:diffpriv} to ensure that the protocol is differentially private and will use secure multi-party computation, which is described in Section~\ref{sec:method:secret_sharing} to achieve cryptographic security. Using these tools, we present our privacy-preserving travel time estimation protocol in Section~\ref{sec:method:alg}

\subsection{Differential Privacy via the Laplace Mechanism}\label{sec:method:diffpriv}

As previously mentioned, our goal is to compute a differentially private approximation to $s(k \Delta t)$ for every $k \in \mathbb{N}$. For this we will use the Laplace Mechanism \cite{DworkMNS06} , which produces a differentially private estimate $S(k \Delta t)$ to $s(k \Delta t)$ based on the following rule 
\begin{align*}
    S(k \Delta t) := s(k \Delta t) + Z 
\end{align*}
where $Z \in \mathbb{R}^{m}$ has independent and identically distributed entries according to the Laplace distribution with mean $0$ and scale parameter $\frac{1}{\epsilon}$ defined below.

\begin{definition}[Laplace Distribution]
The Laplace Distribution with mean $0$ and scale parameter $\frac{1}{\epsilon}$ is a probability distribution over $\mathbb{R}$ denoted as $\cL_{\epsilon}$ with probability density function given by
\begin{align}\label{eqn:laplace_pdf}
    \cL_{\epsilon}(z) := \frac{\epsilon}{2} e^{-\epsilon \abs{z}} \text{ for } z \in \mathbb{R}.
\end{align}
\end{definition}

We use the Laplace mechanism because it provides a differentially private approximation with the minimum possible mean absolute error \cite{Geng14}. 

\begin{fact}
The mechanism $s_e(t) \mapsto S_e(t)$ is $2\epsilon$-differentially private.
\end{fact}

Since we are interested in a decentralized and trustless computation model, the following remark shows that care must be taken in the computation of $S(k \Delta t)$, particularly pertaining to the computation of $Z$, in order for differential privacy to be achieved. 

\begin{remark}[$Z$ must remain hidden]\label{rem:hide_z}
It is essential to the differential privacy of $S(k \Delta t)$ that no observer learns the value of $Z$. Since $S(k \Delta t)$ is announced as the output of the protocol, if in addition $Z$ is known by an observer then that observer can reconstruct $s(k \Delta t)$ by computing $S(k \Delta t) - Z$. In this case, the computation of $S(k \Delta t)$ is not cryptographically secure because the observer learns more about the user data than $S(k \Delta t)$ since in particular it learns the value of $s(k \Delta t)$. 
\end{remark}

In light of Remark~\ref{rem:hide_z}, in the next subsection we discuss a cryptographic technique called secret sharing which we will use for a cryptographically secure computation of $S(k \Delta t)$.  


\subsection{Secure MPC via Secret Sharing}\label{sec:method:secret_sharing}

In this section we review a cryptographic tool known as secret sharing and discuss how different variants of it can be used to enable cryptographically secure arithmetic operations on private data. We describe how cryptographically secure addition can be performed on private data in Section~\ref{sec:method:secret_sharing:smpa} using Additive Secret Sharing, and how cryptographically secure multiplication can be performed on private data in Section~\ref{sec:method:secret_sharing:smpm} using Shamir Secret Sharing. 

\subsubsection{Secure Multi-Party Addition via Additive Secret Sharing}\label{sec:method:secret_sharing:smpa}

Suppose there are $N$ agents $a_1,...,a_N$ and someone wants to share a secret value $x \in \mathbb{N}$ with the agents so that the $N$ agents can reconstruct $x$ if they work together, but no group of fewer than $N$ agents can reconstruct the secret. This can be done using Additive Secret Sharing. 

In Additive Secret Sharing, a large prime integer $p$ is first chosen. The shares $s_1, s_2, ..., s_{N-1}$ are all chosen independently and uniformly at random from the set $\bigbrace{0,1,2,..., p-1}$ and the final share is determined by $s_{N} := x - \sum_{i=1}^{N-1} s_i \mod p$. Finally, $s_i$ is given to $a_i$ for each $1 \leq i \leq N$. 

First, note that the $N$ agents can reconstruct $x$ by simply adding all of their shares together since by construction we have $\sum_{i=1}^N s_i = x$. 

Next note that any group of strictly fewer than $N$ agents cannot reconstruct the secret. A straightforward calculation shows that for any strict subset $S \subset [N]$, the distribution of $\bigbrace{s_i}_{i \in S}$ does not depend on $x$, and therefore $\bigbrace{s_i}_{i \in S}$ provides no information on the value of $x$.  

\begin{example}[Secure Multi-Party Addition]\label{ex:smpa}
One valuable application of Additive Secret Sharing is cryptographically secure computation of the sum of agents' private data. Given $N$ agents $a_1,...,a_N$ with private data $x_1,...,x_N$, their objective is to compute $x := \sum_{i=1}^N x_i$. For each $1 \leq i \leq N$, $a_i$ shares $x_i$ via Additive Secret Sharing by producing shares $s_{1,i}, s_{2,i}, ..., s_{N,i}$ where $s_{j,i}$ is given to $a_j$. At the end of this process, $a_i$ has received $\bigbrace{s_{i,j}}_{j=1}^N$ and can compute $s_i := \sum_{j=1}^N s_{i,j}$. The important observation here is that $\bigbrace{s_i}_{i=1}^N$ are additive secret shares for $x$. Hence the agents can share the values $\bigbrace{s_i}_{i=1}^N$ with one another and compute $x$ via
\begin{align*}
    \sum_{i=1}^N s_i = \sum_{i=1}^N \sum_{j=1}^N s_{i,j} = \sum_{j=1}^N \sum_{i=1}^N s_{i,j} = \sum_{j=1}^N x_j = x. 
\end{align*}
\end{example}

\subsubsection{Secure Multi-Party Multiplication via Shamir Secret Sharing}\label{sec:method:secret_sharing:smpm}

Shamir Secret Sharing \cite{Shamir79} offers a more general $k$-of-$N$ method for secret sharing. In a setting with $N$ agents $a_1,...,a_N$ and a secret $x$ to be shared among them, a $k$-of-$N$ secret sharing scheme assigns shares $s_1,...,s_N$ to the agents so that any subset of $k$ agents can recover $x$, but no subset of fewer than $k$ agents can recover $x$. Note that Additive Secret Sharing is a $N$-of-$N$ scheme. 

Shamir's Secret Sharing is based on the fact that a $k-1$ degree polynomial is uniquely determined from $k$ evaluations. As in the Additive Secret Sharing setting, a large prime $p$ is chosen. To share a secret value $x$, the sharer generates a random $k-1$ degree polynomial
\begin{align*}
    X(z) = x + \sum_{\ell = 1}^{k-1} C_\ell z^\ell
\end{align*}
where $C_1, C_2, ..., C_{k-1}$ are independent and uniformly distributed over $\bigbrace{0,1,...,p-1}$. The share given to $a_i$ is $s_i := X(i) \mod p$. By construction, the shares and coefficients satisfy the following linear relationship
\begin{align*}
    \bigbra{ 
        \begin{tabular}{c}
             $s_1$ \\
             $s_2$ \\
             $\vdots$ \\
             $s_N$
        \end{tabular}
    }
    =
    \bigbra{ 
        \begin{tabular}{c}
             $X(1)$ \\
             $X(2)$ \\
             $\vdots$ \\
             $X(N)$
        \end{tabular}
    }
    = 
    \underbrace{\bigbra{ 
        \begin{tabular}{cccc}
             $1^0$ & $1^2$ & $...$ & $1^{k-1}$\\
             $2^0$ & $2^2$ & $...$ & $2^{k-1}$ \\
             $\vdots$ & $\vdots$ & $\vdots$ & $\vdots$ \\
             $N^0$ & $N^2$ & $...$ & $N^{k-1}$
        \end{tabular}
    }}_{V_{N,k}}
    \bigbra{ 
        \begin{tabular}{c}
             $x$ \\
             $C_1$ \\
             $\vdots$ \\
             $C_{k-1}$
        \end{tabular}
    }
    \mod p
\end{align*}
where $V_{N,k}$ is the $N \times k$ Vandermonde matrix. 

Since $X$ is a degree $k-1$ polynomial, any group of $k$ agents can solve a linear system to obtain the values of $x, C_1, ... C_{k-1}$ and thus recover the secret. 

However, for any subset $S \subset [N]$ with $\abs{S} < k$, a careful calculation shows that the distribution of $\bigbrace{X(i)}_{i \in S}$ does not depend on $x$ and hence $\bigbrace{X(i)}_{i \in S}$ provides no information on the value of $x$.

\begin{example}[Secure Multi-Party Multiplication]\label{ex:smpm}
Shamir Secret Sharing and Additive Secret Sharing can be used to perform cryptographically secure multiplication. Given $N$ agents $a_1,...,a_N$ with additive secret shares $\bigbrace{s_i}_{i=1}^N, \bigbrace{s_i'}_{i=1}^N$ for the values $x,y$ respectively so that $x = \sum_{i=1}^n s_i$ and $y = \sum_{i=1}^n s_i'$, the goal is for the agents to compute the product $xy$ in a cryptographically secure way. 

The computation of $xy$ will require one round of communication. In this communication round, for each $1\leq i \leq N$, $a_i$ performs Shamir secret sharing for its values $s_i$ and $s_i'$. Specifically, it generates two random polynomials $X_i, Y_i$ of degree $\frac{N-1}{2}$ so that $X_i(0) = s_i$ and $Y_i(0) = s_i'$ and then sends $X_i(j), Y_i(j)$ to agent $j$ for each $1 \leq j \leq N$. 

After the communication, $a_i$ obtains $\bigbrace{X_j(i), Y_j(i)}_{i=1}^N$. From these it computes $X(i) := \sum_{j=1}^N X_j(i)$ and $Y(i) := \sum_{j=1}^N Y_j(i)$. Note that $\bigbrace{X(i)}_{i=1}^N$ are Shamir shares for the polynomial $X := \sum_{j=1}^N X_j$ and similarly $\bigbrace{Y(i)}_{i=1}^N$ are Shamir shares for the polynomial $Y := \sum_{j=1}^N Y_j$. Since the polynomials $\bigbrace{X_j}_{j=1}^N, \bigbrace{Y_j}_{j=1}^N$ all have degree at most $\frac{N-1}{2}$, the polynomials $X,Y$ also have degree at most $\frac{N-1}{2}$. Thus if we define the polynomial $H(z) := X(z) Y(z)$, then $H$ has degree at most $N-1$. 

Now $a_i$ computes $X(i) Y(i)$. By definition, $\bigbrace{X(i)Y(i)}_{i=1}^N$ are Shamir shares for the polynomial $H$. Noting that
\begin{align*}
    H(0) = X(0)Y(0) = \bigpar{ \sum_{j=1}^N X_j(0) } \bigpar{ \sum_{j=1}^N Y_j(0) } = \bigpar{ \sum_{j=1}^N s_j } \bigpar{ \sum_{j=1}^N s_j' } = xy, 
\end{align*}
we have a degree $N-1$ polynomial $H$ whose constant term is the desired value $xy$. Furthermore, the $N$ agents know the value of $H$ at $1,2,...,N$ and hence can solve a linear system to obtain the coefficients of $H$ and thus obtain $xy$. 

The Shamir shares $\bigbrace{X(i)Y(i)}_{i=1}^N$ can be converted into Additive shares $\bigbrace{\theta_i}_{i=1}^N$ by setting $\theta_i := \lambda_i X(i)Y(i)$ where $\lambda_i$ is the $(1,i)$ entry of $V_{N,N}^{-1}$. 
\end{example}

\begin{remark}[Honest Majority Regime]
We would like to mention that the Secure Multi-Party Multiplication scheme we describe in this section requires an \textit{honest majority} assumption on user behavior. This means that the scheme requires at least $\frac{N+1}{2}$ users to be fully honest, meaning that they will follow the protocol and will not collude in any way with any other users. The remaining $\frac{N-1}{2}$ users are assumed to be semi-honest. This is a stronger condition than Assumption~\ref{assump:semihonest} which only requires that all users are semi-honest. There exist Secure Multi-Party Multiplication schemes for the semi-honest setting based on Beaver Triples \cite{DamgardKLPSS13}, but the scheme we presented in this section is more computationally efficient, and in practice honest majority may not be an unreasonable assumption. 
\end{remark}

\subsection{Cryptographically secure and Differentially Private Estimation of Travel Times}\label{sec:method:alg}

In this section we show how the differential privacy and secret sharing tools we have discussed can be used to construct a privacy-preserving and decentralized protocol for travel time estimation, which can then be used by a routing service to recommend routes as per Remark~\ref{rem:tt_2_routing}. The protocol is described in Algorithm~\ref{alg:main}. 

In order to satisfy the privacy requirements as stated in Definition~\ref{def:privacy}, our protocol must be both differentially private and cryptographically secure. Recall from Section~\ref{sec:method:diffpriv} that we will use the Laplace mechanism to obtain a differentially private estimate $S(k \Delta t)$ to the traffic counts, which is defined below 
\begin{align*}
S(k \Delta t) = s(k \Delta t) + Z    
\end{align*}
where $Z$ is an i.i.d. vector of $\cL_\epsilon$ distributed random variables. We thus need to compute $S(k \Delta t)$ in a decentralized and cryptographically secure way.

In this section we demonstrate how to compute one entry of the vector $S(k \Delta t)$, i.e., $s_e(k \Delta t) + Z_e$ for an edge $e \in E$. The computation of the entire vector $S(k \Delta t)$ is obtained by parallelizing the computation of all entries. 

We begin by choosing a large prime integer $p$. Inverse transform sampling is a method which can transform a uniform random variable into a random variable with a desired distribution. Using this method we can compute $Z_e = F^{-1}(U_e)$ where $U_e$ is uniformly distributed over $\bigbrace{0,1,...,p-1}$ and $F^{-1}$ is a scaled version of the cumulative distribution function of $\cL_\epsilon$. Concretely, $F^{-1}$ is given by
\begin{align*}
    F^{-1}(u) &= \casewise{
        \begin{tabular}{cc}
            $\frac{1}{\epsilon} \ln \bigpar{ \frac{2u}{p} }$ & for $u \leq \frac{p}{2}$, \\
            $-\frac{1}{\epsilon} \ln \bigpar{ 2 \bigpar{ 1 - \frac{u}{p}} } $ & if $u > \frac{p}{2}$. 
        \end{tabular}
    }
\end{align*}
To make the sampling of $Z_e$ more computationally efficient, we will approximate $F^{-1}$ with a degree $d$ polynomial $P_{\epsilon,d}$. A larger degree leads to a more accurate approximation of the Laplace distribution but comes at a computational cost. 

Thus approximately computing $S_e(k \Delta t)$ amounts to computing $s_e(k \Delta t) + P_{\epsilon,d}(U_e)$ where $U_e$ is required to be uniformly distributed over $\bigbrace{0,1,...,p-1}$. First, additive shares $\bigbrace{\alpha_1,...,\alpha_{N(k \Delta t)}}$ for $s_e(k \Delta t)$ can be computed using Secure Multi-Party Addition as is done in Example~\ref{ex:smpa}. Next, note that if $Y_1, Y_2,..., Y_{N(k \Delta t)}$ are independent and uniformly random on $\bigbrace{0,1,...,p-1}$, then $\sum_{i=1}^{N(k \Delta t)} Y_i \mod p$ is also uniformly distributed. Therefore user $i$ will draw a random value $U_{e,i}$ so that the value $U_e := \sum_{i=1}^{N(k \Delta t)} U_{e,i}$ will be uniformly distributed. Using Secure Multi-Party Addition, the users can obtain additive shares $\bigbrace{ \beta_{1,1}, ... \beta_{1,N(k \Delta t)} }$ for $U_e$. The users will need to compute $P_{\epsilon,d}(U_e)$. Since $P_{\epsilon,d}$ is a polynomial of degree $d$, the users will need additive shares for $U_e^2, U_e^3, ... U_e^d$ for the computation of $P_{\epsilon,d}(U_e)$. The users can obtain such shares through Secure Multi-Party Multiplication by multiplying $U_e$ with itself using Shamir and Additive Secret Sharing as described in Example~\ref{ex:smpm}. Using this method the users obtain additive shares $\bigbrace{\beta_{z,1}, ... \beta_{z,N(k \Delta t)}}_{z=0}^d$ for $\bigbrace{U_e^z}_{z=0}^d$ respectively. Letting $c_0,...,c_d$ be the coefficients of $P_{\epsilon,d}$ so that $P_{\epsilon,d}(u) = \sum_{z = 0}^d c_z u^z$, the users can now construct additive shares for $s_e(k \Delta t) + P_{\epsilon,d}(U_e)$ by taking linear combinations of previously computed shares. Explicitly, the shares
\begin{align*}
    \bigbrace{\alpha_i + \sum_{z=0}^d c_z \beta_{z,i}}_{i=1}^{N(k \Delta t)}
\end{align*}
are additive shares for $s_e(k \Delta t) + P_{\epsilon,d}(U_e)$ since by construction we have
\begin{align*}
    \sum_{i=1}^{N(k \Delta t)} \bigpar{ \alpha_i + \sum_{z=0}^d c_z \beta_{z,i} } &= \bigpar{ \sum_{i=1}^{N(k \Delta t)} \alpha_i} +  \sum_{z=0}^d c_z \bigpar{ \sum_{i=1}^{N(k \Delta t)} \beta_{z,i} } \\
    &= s_e(k \Delta t) + \sum_{z=0}^d c_z U_e^z \\
    &= s_e(k \Delta t) + P_{\epsilon,d}(U_e). 
\end{align*}
Algorithm~\ref{alg:main} describes the procedure that each user performs to enable the decentralized and privacy-preserving computation of $S(k \Delta t)$. 
\begin{algorithm}
\caption{Private and Distributed Traffic Count Estimation}\label{alg:main}
\textbf{Parameters:} Large prime number $p$, approximate inverse CDF $P_{\epsilon,d}(u) := \sum_{z=0}^d c_z u^z$ for the Laplace distribution\;
\textbf{Inputs:} Location information $s(k \Delta t,i)$ for user $i$\;
\textbf{Output:} Estimated traffic counts $S(k \Delta t)$\;
\For{$e \in E$}{
Using Additive Secret Sharing, obtain a share $\alpha_i$ of $s(k \Delta t) = \sum_{j=1}^{N(k \Delta t)} s(k \Delta t, j)$ through Secure Multi-Party Addition\;
Draw $U_{e,i}$ uniformly at random over $\bigbrace{0,1,...,p-1}$\;
Using Additive Secret Sharing, obtain a share $\beta_{1,i}$ of $U_e := \sum_{j=1}^{N(k \Delta t)} U_{e,j}$ through Secure Multi-Party Addition\;
    \For{$1 \leq z \leq d$}{
        Using Shamir and Additive Secret Sharing, obtain a share $\beta_{z,i}$ of $U_e^z$ through Secure Multi-Party multiplication\;
    }
Compute $\theta_i := \alpha_i + \sum_{z=0}^d c_z \beta_{z,i}$ and send $\theta_i$ to all other users\;
$S_e(k \Delta t) \leftarrow \sum_{j=1}^{N(k \Delta t)} \theta_j$\;
}
\textbf{Return} $S(k \Delta t)$\;
\end{algorithm}


\section{Analysis: Accuracy of travel times based on $S(k \Delta t)$}\label{sec:theory}

In the previous section we presented a decentralized and privacy-preserving protocol for computing $S(k \Delta t)$, which is a differentially private estimate of the traffic counts $s(k \Delta t)$ at time $k \Delta t$ (i.e., the $k$th timestep). Since the traffic counts are useful for travel time estimation through volume delay functions, one natural question is how the travel time estimates obtained from $S(k \Delta t)$ will differ from those obtained from the ground truth traffic counts $s(k \Delta t)$. In this section we show that if the roads in the transportation network $G$ are sufficiently large, then the travel time estimates obtained from $S(k \Delta t)$ will be close to those obtained had we used the non-privacy-preserving ground truth value $s(k \Delta t)$.  

We first discuss the errors in estimating the traffic counts due to the Laplace mechanism in Section~\ref{sec:analysis:laplace_accuracy}. Next, in Section~\ref{sec:analysis:protocol_acc} we show how the properties of volume delay functions mitigate the errors induced by the Laplace mechanism, and how composing these two together can help achieve accurate and private travel time estimates.

\subsection{Accuracy of the Laplace Mechanism}\label{sec:analysis:laplace_accuracy}

Recall that $S_e(k \Delta t) = s_e(k \Delta t) + Z_e$ is a differentially private estimate of the traffic count on road $e$ at time $t$. 
The mean absolute percentage error (MAPE) of $S_e(k \Delta t)$ as an estimate for $s_e(k \Delta t)$ is given by
\begin{align*}
    \frac{\mathbb{E} \bigbra{ \abs{S_e(k \Delta t) - s_e(k \Delta t)}}}{s_e(k \Delta t)} &= \frac{\mathbb{E} \bigbra{\abs{Z_e}}}{s_e(k \Delta t)} \\
    &= \frac{1}{\epsilon s_e(k \Delta t)}.
\end{align*}

From this we can make two conclusions. When there is a lot of traffic on $e$, meaning that $s_e(k \Delta t)$ is much larger than $\frac{1}{\epsilon}$, then $\epsilon s_e(k \Delta t)$ will be large and hence $S_e(k \Delta t)$ will have a small MAPE. However, if $s_e(k \Delta t)$ is small, then $S_e(k \Delta t)$ will have a large MAPE. 

This shows us that the Laplace Mechanism has poor accuracy when reporting small values. In fact, this is true for all differentially private mechanisms since the Laplace Mechanism has the minimum mean absolute error among all differentially private mechanisms \cite{Geng14}. This observation is consistent with Example~\ref{ex:cryptosec_not_enough} in that sparse data and small values pose the most difficulty in privacy-preserving efforts. 

Fortunately, this bad news does not end our hopes for achieving both accuracy and privacy in travel time estimation. Even if $S_e(k \Delta t)$ may not always be a good estimate for $s_e(k \Delta t)$, recall that our ultimate objective is travel time estimation, so we are interested in how well $\tau_e(S_e(k \Delta t))$ approximates $\tau_e(s_e(k \Delta t))$. Recall Definition~\ref{def:counts2time} for a description of $\tau_e$. Next we will show how properties of delay functions can enable accurate travel time estimates even if traffic count estimation is poor. 

\subsection{Protocol accuracy for travel time estimation}\label{sec:analysis:protocol_acc}

In this section we show that if a road $e \in E$ is sufficiently large, then the travel time estimates computed from $S_e(k \Delta t)$ are close to the travel time estimates computed from the ground truth $s_e(k \Delta t)$. Mathematically, this means that $\tau_e(S_e(k \Delta t))$ is a good estimate for $\tau_e(s_e(k \Delta t))$. 

The key insight behind our result lies in the complementary qualities of differential privacy and volume delay functions. As we saw in Section~\ref{sec:analysis:laplace_accuracy}, the Laplace mechanism has good accuracy when reporting large values, but poor accuracy for reporting small values. Volume delay functions on the other hand, are very sensitive when the input is large, but very insensitive when the input is small. When composing a volume delay function with a Laplace mechanism, the complementary qualities manifest in two ways. When the traffic $s_e(k \Delta t)$ is larger than a road's capacity, the volume delay function is very sensitive. Fortunately, in this case the high accuracy of the Laplace mechanism ensures that the traffic count is estimated accurately, leading to accurate traffic flow estimation, which leads to accurate travel time estimation. On the other hand, when the traffic is below the road's capacity, the Laplace mechanism has poor accuracy, however the delay function is very insensitive in this regime and is able to tolerate large estimation error, leading to accurate travel time estimation. Thus the Laplace mechanism and volume delay function cover each others' weaknesses to enable accurate travel time estimation for any level of traffic. 

We formalize this insight through Theorem~\ref{thm:tt_accuracy}, which, given desired privacy and accuracy levels $\epsilon$ and $\delta$ respectively, provides conditions under which $\tau_e(S_e(k \Delta t))$ will be close to $\tau_e(s_e(k \Delta t))$ with high probability. The condition is determined by the road's $\delta$-critical traffic count, which is defined below.

\begin{definition}[$\delta$-critical traffic count]
The $\delta$-critical traffic count of a road $e \in E$ is the number of vehicles on the road in steady state so that the travel time is exactly $1+\delta$ times as large as its free-flow travel time. Mathematically, the $\delta$-critical capacity of $e$ is
\begin{align*}
    F_e(c_{e,\delta}) = c_{e,\delta} f_e(c_{e,\delta}) = (1+\delta) c_{e,\delta} f_e(0)
\end{align*}
where $c_{e,\delta}$ is the $\delta$-capacity of the road $e$ as defined in Defintion~\ref{def:delta_cap}. With this setup in place, we now present Theorem~\ref{thm:tt_accuracy}. 
\end{definition}

\begin{theorem}[Accuracy of Travel Time Estimates]\label{thm:tt_accuracy}
Let $\epsilon, \delta \geq 0$ specify the desired privacy and accuracy levels respectively and let $p \in [0,1]$ represent a failure probability. For a road $e \in E$, if $f_e$ satisfies Assumption~\ref{assump:traveltime} and 
\begin{align*}
    (1+\delta) c_{e,\delta} f_e(0) \geq \frac{1}{\epsilon} \bigpar{ \frac{1}{\delta} + 1 } \log \frac{1}{p}
\end{align*}
where $c_{e,\delta}$ is the $\delta$-capacity of $e$, then for any value of $s_e(k \Delta t) \in \mathbb{N}$, the following condition is satisfied with probability at least $1-p$:
\begin{align*}
    \frac{\abs{ \tau_e \bigpar{S_e(k \Delta t)} - \tau_e \bigpar{s_e(k \Delta t)} }}{\tau_e \bigpar{s_e(k \Delta t)}} \leq \delta. 
\end{align*}
\end{theorem}

\noindent See Appendix~\ref{pf:thm:tt_accuracy} for a proof of Theorem~\ref{thm:tt_accuracy}. Next, we will discuss whether the condition required by Theorem~\ref{thm:tt_accuracy} is satisfied in practice by roads in real transit networks.

\subsection{Discussion}\label{sec:theory:disc}

One natural and immediate question is whether the requirement on the $\delta$-critical traffic count of roads in Theorems \ref{thm:tt_accuracy} are satisfied by real road networks. To answer this question we first discuss parameter choices. To ensure a meaningful privacy guarantee for each timestep, $\epsilon$ should be significantly smaller than $1$. For this reason we focus on applications where $\epsilon = 0.2$. 

\begin{figure}
    \centering
    \includegraphics[width = 0.5\linewidth]{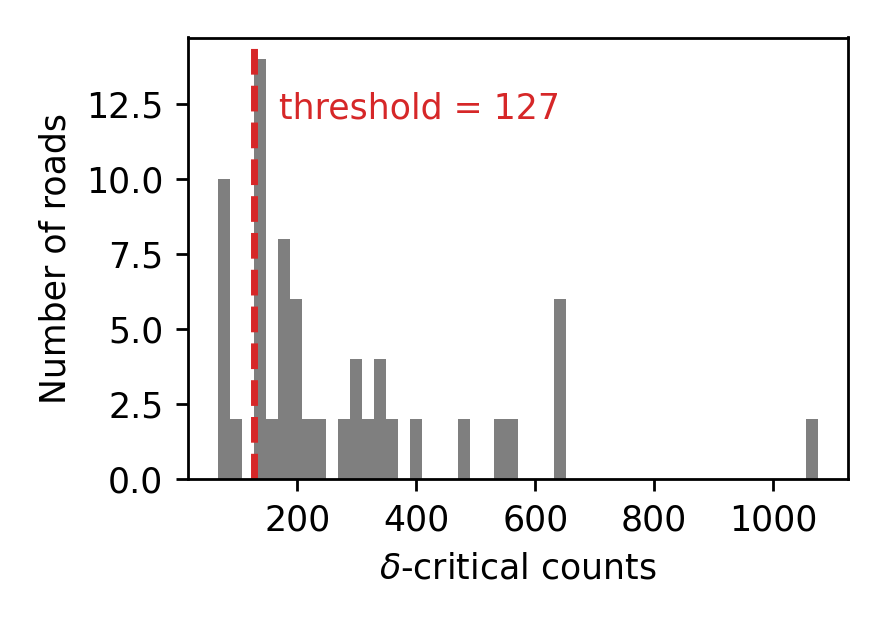}
    \caption{Distribution of $\delta$-critical counts for all roads in the network for $\delta = 0.1$. The dashed red line denotes the threshold above which the road travel time estimate will have an error less than 10\%, 90\% of the time to give a privacy level of $\epsilon=0.2$.}
    \label{fig:realistic_capacity}
\end{figure}

With $\epsilon = 0.2$, $\delta = 0.1$ and $p = 0.1$, the condition in Theorem~\ref{thm:tt_accuracy} requires that a road's $\delta$-critical count be at least $127$ cars. For such roads, the Theorem states that the estimated travel time will be within 10 percent of the ground truth with probability at least 90 percent. To see whether such a requirement is reasonable, we studied a real world transportation network. The $\delta$-critical capacities of all roads in the Sioux Falls transportation network are plotted in Figure~\ref{fig:realistic_capacity}. The figure shows that more than 80 percent of the roads in the Sioux Falls network have $\delta$-critical counts above 127. Thus the conditions required by Theorem~\ref{thm:tt_accuracy} are realistic for most roads. 

Care must also be taken in choosing the value of $\Delta t$, i.e. the amount of time between updates to the estimated traffic counts. Specifically, $\Delta t$ should be chosen so that it is similar to the typical travel time of a road in the network. If $\Delta t$ is much smaller than the typical travel time on a road, then sample average approximation can be used to denoise $S(k \Delta t)$ to get a better estimate of $s(k \Delta t)$. While better estimation is usually good, it is also synonymous with less privacy. To illustrate this concept, suppose that for timesteps $1, 2, 3, ..., N$ the ground truth traffic counts for a given road $e$ is constant, meaning that there is a value $s_e$ so that $s_e(k \Delta t) = s_e$ for all $1 \leq k \leq N$. Now for each $k$, $S_e(k \Delta t)$ is an unbiased estimator for $s_e$ with variance $\frac{2}{\epsilon^2}$. This variance is necessary to ensure differential privacy. However, note that $\frac{1}{N} \sum_{k'=1}^N S_e(k' \Delta t)$ is also an unbiased estimator for $s_e$ but now has variance $\frac{2}{N \epsilon^2}$. This decrease in variance leads to worse privacy guarantees. For this reason, $\Delta t$ should be chosen similar to the travel time of a road so that $N$ can never get too large. 

\section{Numerical Experiments}\label{sec:experiments}

We evaluate the performance of our protocol in the Sioux Falls road network setting. The purpose of these numerical experiments is to compare the travel time experienced by vehicles when they are routed  using (a) travel time estimates derived from our protocol and (b) travel time estimates derived from the ground truth traffic counts. In particular we want to quantify the extent of travel time degradation incurred by the use of our protocol's privacy-preserving mechanisms. Thus, these experiments help us test the practical implications of our road-level theoretical results when viewed in the context of the entire network. In Section \ref{sec:experiments:setup}, we present details about the data sources, road network, traffic demand, and the experimental setup. In Section \ref{sec:experiments:result}, we study the impact of our protocol on route choices and travel time. These simulations suggest that the price for privacy may be negligible or even zero, thereby strengthening the case for conducting real-world field studies.

\subsection{Setup}\label{sec:experiments:setup}

The properties of the Sioux Falls road network as well as the typical user demands is obtained from the \href{https://github.com/bstabler/TransportationNetworks}{Transportation Network Test Problems (TNTP)} dataset. The Sioux Falls road network consists of 24 nodes and 76 edges. Each edge is characterized by a maximum speed, free flow throughput (i.e., vehicles per hour), and length of the segment. Note that we use the terms edge and road interchangeability. Travel times are computed using BPR functions whose parameters are obtained from the aforementioned edge characteristics. Additionally, the dataset also reports the steady state traffic demand between 528 origin-destination (OD) pairs. 

We conduct two types of experiments: private routing and non-private routing. In both types of experiments we simulate traffic flow on the network at a time resolution of $10$ seconds. At every time step, we draw new demand from a Poisson distribution, with the mean demand proportional to the steady state demand reported in the dataset. Thus, at each time step, we draw a random number of vehicles with a corresponding origin and destination. For each vehicle, we identify a shortest travel time route between its origin and destination nodes. In the private routing experiment, the shortest path is computed using the most recent travel time estimates produced by our protocol. In the non-private routing experiment, the shortest path is computed using the ground truth travel time which depends on the ground truth traffic counts. In both types of experiments, the simulated movement of the vehicles on the road network is determined by the ground truth traffic counts on that road.


The duration of the simulation is 2 hours, with additional buffer time in the end for vehicles already in transit to complete their trips. For the private routing experiments, the vehicles are assumed to use our protocol to update travel time estimates every $\Delta t = 2$ minutes to minimize the number of reports that a vehicle makes from the same road (See Section~\ref{sec:theory:disc}). In our experiments, we consider $\epsilon$ values of $0.01$ and $0.1$. We will discuss how our protocol would perform for other values of $\epsilon$ in Section \ref{sec:experiments:result}. We consider three vehicle demand profiles. In the baseline scenario, about 60,120 vehicles are expected to join the road network every hour. We also consider a low demand scenario, where all the OD Poisson parameters are decreased by 50\% and a high demand scenario, where all the OD Poisson parameters are increased by 50\%. To gain some more insight into the degree of strain this demand induces on the road network, we refer the reader to Table \ref{tab:demand}. Here, we report the rate of vehicles being added to the network for each scenario as well as the minimum, maximum, and average road utilization during the simulation period. The road utilization is defined as the term $x/c$ from Remark \ref{rem:bpr}. Utilization greater than 1 indicates congestion on a road. The baseline demand profile results in several congested roads -- representing a realistic scenario for testing our protocol. Additionally, as expected, increasing the demand rate results in a higher road utilization and more congested roads.

\begin{table}[]
\begin{center}
\begin{tabular}{c|c|c|c|c}
\textbf{Scenarios} & \textbf{\begin{tabular}[c]{@{}c@{}} Rate \\ (vehicles/hr)\end{tabular}} & \textbf{\begin{tabular}[c]{@{}c@{}}Min road \\ utilization\end{tabular}} & \textbf{\begin{tabular}[c]{@{}c@{}}Max road \\ utilization\end{tabular}} & \textbf{\begin{tabular}[c]{@{}c@{}}Average road \\ utilization\end{tabular}} \\ \hline \hline
Baseline        & 60,100                                                                        & 0.05                                                                     & 1.15                                                                     & 0.52                                                                         \\ \hline
Low             & 30,050                                                                        & 0.00                                                                     & 0.90                                                                      & 0.28                                                                         \\ \hline
High            & 90,150                                                                        & 0.07                                                                     & 1.47                                                                     & 0.74                                                                        
\end{tabular}
\end{center}
\caption{Demand scenarios used in our simulation}
\label{tab:demand}
\end{table}

\subsection{Results}\label{sec:experiments:result}


We evaluate the impact of privacy noise on the routes and travel times of vehicles in the network. Table \ref{tab:eps0.01} presents several performance measures for $\epsilon=0.01$ under the three demand profiles. Note that the choice of $\epsilon = 0.01$ is overly conservative since smaller $\epsilon$ represents a stricter privacy requirement and thus more noise injected into the system. Typical values of $\epsilon$ that are chosen in practice are larger than 0.1 depending on the application. Nevertheless, we consider this relatively stringent privacy requirement to understand the performance limits our protocol. 

From Table \ref{tab:eps0.01}, we first observe that the increase in average travel time for a vehicle is only 8 seconds in the baseline case. This corresponds to a 1.3\% increase in travel time. For the low and high demand scenarios, the increase in average travel times are 3.1 sec (0.6\%) and 13.2 sec (1.9\%) respectively. Vehicles will experience a low additional travel time if the routing decisions for vehicles from our protocol closely resembles what they would have done without any privacy noise. In other words, if the shortest paths on the graph with noisy travel time estimates is the same (or very similar) to the shortest path on a graph with accurate travel time estimates, the vehicles will experience very little additional travel time. Our results confirm that this is indeed the case -- the routes chosen by almost 90\% of the vehicles are unchanged due to our protocol. Finally, as we compare the three demand scenarios, we observe that higher demand results in greater congestion and subsequently higher travel times. Furthermore, when demand is higher, the choice of an appropriate route for each vehicle is even more critical to minimize. Working with noisy estimates of travel time, during periods of high congestion can thus lead to incorrect routing choices. This is consistent with our observation that the route of 90.9\% of the cars are unchanged by our protocol when the demand is low, but only 87\% of the routes remain unchanged when the demand is high.

\begin{table}[]
\begin{tabular}{c|c|c|c}
\textbf{Performance measure}                      & \textbf{Low demand} & \textbf{Baseline demand} & \textbf{High demand} \\ \hline \hline
Travel time (sec)                         & 546.5               & 591.9                    & 678.2                \\ \hline
Travel time with our protocol (sec)       & 549.6               & 599.9                    & 691.4                \\ \hline
Increase in travel time (sec)             & 3.1                 & 8.0                      & 13.2                 \\ \hline
Increase in travel time (\%)              & 0.6                 & 1.3                      & 1.9                  \\ \hline
Cars with no change in route (\%)         & 90.9                & 88.3                     & 87.1                 \\ \hline
Cars with no increase in travel time (\%) & 65.9                & 41.3                     & 20.6                
\end{tabular}
\caption{Performance measures for $\epsilon = 0.01$}
\label{tab:eps0.01}
\end{table}

\begin{table}[]
\begin{tabular}{c|c|c|c}
\textbf{Performance measure}                      & \textbf{Low demand} & \textbf{Baseline demand} & \textbf{High demand} \\ \hline \hline 
Travel time (sec)                         & 546.5               & 591.9                    & 678.2                \\ \hline
Travel time with our protocol (sec)       & 546.5               & 592.2                    & 677.4                \\ \hline
Increase in travel time (sec)             & 0.0                & 0.2                     & -0.7                 \\ \hline
Increase in travel time (\%)              & 0.0                   & 0.0                     & -0.1                 \\ \hline
Cars with no change in route (\%)         & 98.4                & 97.5                     & 94.4                 \\ \hline
Cars with no increase in travel time (\%) & 90.7                & 67.9                     & 38.6                
\end{tabular}
\caption{Performance measures for $\epsilon = 0.1$}
\label{tab:eps0.1}
\end{table}

Next, we study the performance of our protocol with a lower privacy setting of $\epsilon=0.1$. Note that we expect that the performance of our protocol converges to the non-private setting as $\epsilon \rightarrow \infty$. Interestingly, our results show that even with $\epsilon=0.1$, the performance of our protocol becomes indistinguishable to the non-private setting. The performance of our protocol for $\epsilon=0.1$ is shown in Table \ref{tab:eps0.1}. We observe that the increase in travel time is nearly 0 for all the three demand scenarios. In fact, the randomness in the travel time estimates can also result in marginal improvements in travel time in some settings (reflected as a negative increase in travel time). Not surprisingly, the routes chosen by nearly all the cars also is unaffected by our privacy preserving protocol. The results from varying $\epsilon$ are very encouraging -- for a reasonable privacy requirement, we are able to get privacy for `free' with no loss in system performance. Although not shown for brevity, our experiments indicated that cars observe no increase in average travel time for all values of $\epsilon>0.1$.

The results from these experiments are significantly better than what is guaranteed by Theorem~\ref{thm:tt_accuracy}. As we discussed in Section~\ref{sec:theory:disc}, Theorem~\ref{thm:tt_accuracy} promises that the estimated travel time on each road will be within 10 percent of the ground truth at least 90 percent of the time when $\epsilon = 0.2$. Our numerical results in Table~\ref{tab:eps0.01} show that even when $\epsilon = 0.01$ (i.e., $20$ times as noisy as $\epsilon = 0.2$), our protocol only introduces an overhead of 8 percent in the baseline case and 13 percent in a high demand case. We believe this is because real world road networks have redundancy. By this we mean that there are many near-optimal paths from an origin to a destination, so it is very likely that at least one of these paths has an accurately estimated travel time. Theorem~\ref{thm:tt_accuracy} looks only at the edge level and is thus does not exploit favorable network topologies. On a positive note, Theorem~\ref{thm:tt_accuracy} is general in the sense that it can be applied to a network with any topology and any demand structure. 

\section{Conclusion}\label{sec:conclusion}


In this paper we propose a protocol for a decentralized routing service where travel times are computed from user location data in a privacy-preserving way. In most current routing services, users give their individual location data directly to routing services in exchange for route recommendations. Since this data is associated with the users' identity, users' schedules, habits, preferences and other private information can be inferred through repeated interactions with routing services. Contrary to this, the protocol proposed in this paper is both differentially private and cryptographically secure, meaning that only the aggregate effect of traffic on travel time is obtainable from the protocol, and users' individual location data cannot be inferred by other parties. We also show that for large roads, it is possible to estimate travel time both accurately and privately. This is due to complementary qualities of differential privacy and delay functions. We evaluated the performance of the protocol through simulation in the Sioux Fall transportation network and showed that the protocol incurs minimal performance overhead in practice while providing a principled privacy guarantee.

There are many interesting and important directions for future work. The first direction is related to finding a more refined definition for privacy. Travel time estimation in the literature is often based on flow or average speed of vehicles on the road. However, we chose to estimate travel times based on traffic counts due to compatibility with differential privacy. More specifically, without additional domain-specific assumptions, it is impossible to compute flow or average speed in both an accurate and differentially private way (see Remark~\ref{rem:diffpriv:sensitivity}). Thus while differential privacy is a general and powerful concept, it is perhaps too restrictive for some common mobility applications such as flow or speed estimation. Developing a more specialized notion of privacy for mobility applications could enable more algorithmic possibilities while retaining meaningful privacy guarantees. The second direction is related to adoption rate. In this paper we implicitly assume that all vehicles in the network are willing to participate in the protocol, though technically a uniform and known adoption rate would be sufficient. While we believe this is a reasonable assumption in an era of connected vehicles, developing a protocol that is agnostic to participation rate would provide robustness. A third direction is related to other applications. Developing decentralized and privacy-preserving pricing for roads and for mobility services would be an interesting direction. 

\bibliographystyle{ieeetr}
\bibliography{peripherals/TYGP.bib}

\newpage 

\appendix 

\section{Proof of Theorem~\ref{thm:tt_accuracy}}\label{pf:thm:tt_accuracy}

We prove Theorem~\ref{thm:tt_accuracy} by proving the following Lemma:

\begin{lemma}\label{lem:tt_accuracy}
For any $s \in \mathbb{R}_+$, the following inequality 
\begin{align*}
    \frac{\abs{\tau_e(s+Z) - \tau_e(s)}}{\tau_e(s)} \leq \delta
\end{align*}
is satisfied with probability at least $1-p$ when $Z \sim \cL_\epsilon$.
\end{lemma}
The Theorem follows by applying the result with $s = s_e(k \Delta t)$. We will prove Lemma!\ref{lem:tt_accuracy} by considering two exhaustive cases:
\begin{itemize}
    \item[] Case 1: $s < \frac{\log \frac{1}{p}}{\epsilon \delta}$,
    \item[] Case 2: $s \geq \frac{\log \frac{1}{p}}{\epsilon \delta}$.
\end{itemize}

\subsection{Proving Lemma~\ref{lem:tt_accuracy} in Case 1}

In Case $1$ we have $s < \frac{\log \frac{1}{p}}{\epsilon \delta}$. By the condition in Theorem~\ref{thm:tt_accuracy}, $F_e(c_{e,\delta}) \geq \frac{1}{\epsilon} \bigpar{ \frac{1}{\delta} + 1 } \log \frac{1}{p}$, and thus $s \leq F_e(c_{e,\delta})$. 

Next, note that
\begin{align*}
    \mathbb{P} \bigbra{ s+Z \geq F_e(c_{e,\delta}) } &= \mathbb{P} \bigbra{ Z \geq F_e(c_{e,\delta}) - s } \\
    &\leq \mathbb{P} \bigbra{ Z \geq \bigpar{ \frac{1}{\epsilon} \bigpar{ \frac{1}{\delta} + 1} \log \frac{1}{p} } - \frac{\log \frac{1}{p}}{\epsilon \delta} } \\
    &= \mathbb{P} \bigbra{ Z \geq \frac{\log \frac{1}{p}}{\epsilon} } \\
    &\overset{(a)}{=} \frac{1}{2} \exp \bigpar{ - \epsilon \frac{\log \frac{1}{p}}{\epsilon} } \\
    &= \frac{p}{2},
\end{align*}
where $(a)$ is due to the the formula for the cumulative distribution function for the Laplace distribution. Therefore, with probability at least $1 - \frac{p}{2}$, both $s, s+Z$ are less than $F_e(c_{e,\delta})$. For the remainder of the Case 1 discussion we will condition on the high probability event that both $s, s+Z$ are less than $F_e(c_{e,\delta})$. 

By Assumption~\ref{assump:traveltime} we know that $f_e$ is non-decreasing, which means that $F_e$ is non-decreasing and invertible. From this we can conclude that $F_e^{-1}(s) \leq c_{e,\delta}$ and $F_e^{-1}(s+Z) \leq c_{e,\delta}$. Since $f_e$ is non-decreasing this means
\begin{align*}
    f_e(0) &\leq f_e(s), f_e(s+Z) \leq f_e(c_{e,\delta}).
\end{align*}
From this we can deduce that
\begin{align*}
    \abs{ f_e(s+Z) - f_e(s) } &\leq f_e(c_{e,\delta}) - f_e(0) \\
    &= (1+\delta) f_e(0) - f_e(0) \\
    &= \delta f_e(0) \\
    &\leq \delta f_e(s),
\end{align*}
which establishes Lemma~\ref{lem:tt_accuracy} in Case 1. 

\subsection{Proving Lemma~\ref{lem:tt_accuracy} in Case 2}

For Case 2 we have $s \geq \frac{\log \frac{1}{p}}{\epsilon \delta}$. Our analysis for this case will involve $\frac{d}{dy} F_e^{-1}(y)$ and $\frac{d}{dy} \tau_e(y)$, so we will compute them using chain rule:
\begin{align*}
    1 &= \frac{d}{dy} y \\
    &= \frac{d}{dy} F_e(F_e^{-1}(y)) \\
    &= F_e'(F_e^{-1}(y)) \bigpar{ \frac{d}{dy} F_e^{-1}(y) } \\
    \implies \frac{d}{dy} F_e^{-1}(y) &= \frac{1}{F_e'(F_e^{-1}(y))}.
\end{align*}
Using this, we can now compute $\frac{d}{dy} \tau_e(y)$ using chain rule:
\begin{align*}
    \frac{d}{dy} \tau_e(y) &= \frac{d}{dy} f_e(F_e^{-1}(y)) \\
    &= f_e'(F_e^{-1}(y)) \bigpar{ \frac{d}{dy} F_e^{-1}(y) } \\
    &= \frac{f_e'(F_e^{-1}(y))}{F_e'(F_e^{-1}(y))}
\end{align*}
Defining $x_y := F_e^{-1}(y)$, we see that
\begin{align*}
    \frac{d}{dy} \tau_e(y) &= \frac{f_e'(x_y)}{F_e'(x_y)} = \frac{f_e'(x_y)}{x_y f_e'(x_y) + f_e(x_y)}. 
\end{align*}
We make an observation that will be useful later:
\begin{observation}\label{obs:tau_prime_bound}
Since $f_e$ is non-negative under Assumption~\ref{assump:traveltime}, we have $\frac{d}{dy} \tau_e(y) \leq \frac{1}{x_y}$.
\end{observation}

\begin{observation}\label{obs:t_is_y_over_x}
$t_e(y) = \frac{y}{x_y}$. This is because since $t_e(y) = f_e(F_e^{-1}(y))$, we have $F_e^{-1}(y) t_e(y) = F_e^{-1}(y) f_e(F_e^{-1}(y)) = F_e(F_e^{-1}(y)) = y$. Therefore $t_e(y) = \frac{y}{F_e^{-1}(y)} = \frac{y}{x_y}$. 
\end{observation}

\noindent With this setup in hand, we are now ready to prove the Lemma. First note that
\begin{align*}
    \mathbb{P}\bigbra{ \abs{Z} \geq \delta s } &\overset{(a)}{=} \exp \bigpar{ - \epsilon \delta s } \\
    &\leq \exp \bigpar{ - \epsilon \delta \frac{\log \frac{1}{p}}{\epsilon \delta} } \\
    &= p,
\end{align*}
where $(a)$ is due to the fact that $\abs{Z}$ has the exponential distribution with parameter $\frac{1}{\epsilon}$, and this distribution has the cumulative distribution function $\mathbb{P}[\abs{Z} \geq t] = e^{-\epsilon t} \mathds{1}[t \geq 0]$. Thus with probability at least $1-p$, we have $\abs{Z} \leq \delta s$. For the remainder of the Case 2 discussion we will condition on the high probability event that $\abs{Z} \leq \delta s$. 

By the fundamental theorem of calculus,
\begin{align*}
    \abs{ \tau_e(s+Z) - \tau_e(s) } &= \abs{ \int_{s}^{s+Z} \bigpar{ \frac{d}{dy} \tau_e(y) } dy } \\
    &\leq \int_{s}^{s+Z} \abs{ \frac{d}{dy} \tau_e(y) } dy \\
    &\overset{(a)}{\leq} \int_{s}^{s+Z} \frac{1}{\abs{x_y}}  dy \\
    &\overset{(b)}{\leq} \int_{s}^{s+Z} \frac{1}{x_{\min(s,s+Z)}}  dy \\
    &\overset{(c)}{\leq}\int_{s}^{s+Z} \frac{1}{x_{(1-\delta)s}}  dy \\
    &= \frac{\abs{Z}}{x_{(1-\delta)s}} \\
    &\leq \frac{\delta s}{x_{(1-\delta)s}} \\
    &= \frac{\delta}{1-\delta} \frac{(1-\delta)s}{x_{(1-\delta)s}} \\
    &\overset{(d)}{=} \frac{\delta}{1-\delta} \tau_e((1-\delta)s) \\
    &\overset{(e)}{\leq} \frac{\delta}{1-\delta} \tau_e(s)
\end{align*}
where $(a)$ is due to Observation~\ref{obs:tau_prime_bound}. Since $x_y$ was defined to be $F_e^{-1}(y)$, $(b)$ is due to the fact that $F_e$ is increasing, therefore $x_y$ is an increasing function of $y$ and hence $\frac{1}{x_y}$ is a decreasing function of $y$. $(c)$ is because $\min(s,s+Z) \geq (1-\delta) s$ since we are in the event that $\abs{Z} \leq \delta s$. $(d)$ is due to Observation~\ref{obs:t_is_y_over_x}, and $(e)$ is because $t_e$ is an increasing function, since it is $f_e$ composed with $F_e^{-1}$, which are both increasing. Since $\frac{\delta}{1-\delta} = \delta + \frac{\delta^2}{1-\delta}$, we have
\begin{align*}
    \abs{ \tau_e(s+Z) - \tau_e(s) } &\leq \bigpar{ \delta + O(\delta^2) } \tau_e(s)
\end{align*}
which proves Lemma~\ref{lem:tt_accuracy} in Case 2. 
\end{document}